\documentclass[12pt,eqsecnum,preprint]{aastex}
\title{3-D Simulations of Ergospheric Disk Driven Poynting Jets}
\begin{document}
\author{Brian Punsly}
\affil{4014 Emerald Street No.116, Torrance CA, USA 90503 and
International Center for Relativistic Astrophysics,
I.C.R.A.,University of Rome La Sapienza, I-00185 Roma, Italy}
\email{brian.m.punsly@L-3com.com or brian.punsly@gte.net}
\begin{abstract}This Letter reports on 3-dimensional simulations
of Kerr black hole magnetospheres that obey the general relativistic
equations of perfect magnetohydrodynamics (MHD). In particular, we
study powerful Poynting flux dominated jets that are driven from
dense gas in the equatorial plane in the ergosphere. The physics of
which has been previously studied in the simplified limit of an
ergopsheric disk. For high spin black holes, $a/M > 0.95$, the
ergospheric disk is prominent in the 3-D simulations and is
responsible for greatly enhanced Poynting flux emission. Any large
scale poloidal magnetic flux that is trapped in the equatorial
region leads to an enormous release of electromagnetic energy that
dwarfs the jet energy produced by magnetic flux threading the event
horizon. The implication is that magnetic flux threading the
equatorial plane of the ergosphere is a likely prerequisite for the
central engine of powerful FRII quasars.
\end{abstract}
\section{Introduction}Recent studies of luminous radio quasars indicate that the power of the radio jet
can exceed the bolometric luminosity associated with the accretion
flow thermal emission \citep{pun06,pun07}. This has proven to be
quite challenging for current 3-D numerical simulations of MHD black
hole magnetospheres. Based on table 4 of \citet{haw06} and the
related discussion of \citet{pun06,pun07}, the most promising 3-D
simulations for achieving this level of efficiency are those of the
highest spin, $a/M \approx 1$ (where the black hole mass, $M$, and
the angular momentum per unit mass, $a$, are in geometrized units).
More generally, such high spins have been inferred in some black
hole systems based on observational constraints \citep{mcc06}. Thus,
there is tremendous astronomical relevance to these highest spin
configurations, in particular the physical origin of the
relativistic Poynting jet. The first generation of long term 3-D
simulations produced one Poynting flux powerhouse, the $a/M=0.995$
simulation, KDE \citep{dev03,dev05,hir04,kro05}. The source of most
of the Poynting flux was clearly shown to be outside the event
horizon in KDE \citep{pun05}. However, without access to the
original data, the details of the physical mechanism could not be
ascertained. A second generation of 3-D simulations were developed
in \citet{haw06}, the highest spin case was KDJ, $a/M=0.99$, with by
far the most powerful Poynting jet within the new family of
simulations; three times the Poynting flux (in units of the
accretion rate of mass energy) of the next closest simulation KDH,
$a/M=0.95$. The last three data dumps, at simulation times, t = 9840
M, t = 9920 M and t = 10000 M, were generously made available to
this author. The late time behavior of the simulations is
established after t = 2000 M (when the large transients due to the
funnel formation have died off) making these data dumps of
particular interest for studying the Poynting jet \citep{haw06}.
This paper studies the origin of the Poynting jet at these late
times.
\par The analysis of the data from the KDJ simulation clearly indicates
that the Poynting flux in the outgoing jet is dominated by large
flares. Typically, one expects the turbulence in the field variables
to mask the dynamics of Poynting flux creation in an individual time
slice of one of the 3-D simulations \citep{pun05}. Surprisingly, the
flares are of such a large magnitude that they clearly standout
above the background field fluctuations as evidenced by figure 1.
The flares are created in the equatorial accretion flow deep in the
egosphere between the inner calculational boundary at r=1.203 M and
r= 1.6 M (the event horizon is at r= 1.141 M). Powerful beams of
Poynting flux emerge perpendicular to the equatorial plane in the
ergospheric flares and much of the energy flux is diverted outward
along approximately radial trajectories that are closely aligned
with the poloidal magnetic field direction in the jet (see figure
1). The situation is unsteady, whenever some vertical magnetic flux
is captured in the accretion flow it tends to be asymetrically
distributed and concentrated in either the northern or southern
hemisphere. This hemisphere then receives a huge injection of
electromagnetic energy on time scales $\sim 60 M $.
\par The source of Poynting flux in KDJ resembles a nonstationary
version of the ergospheric disk (see \citet{pun90} and chapter 8 of
\citet{pun01} for a review). The ergospheric disk is modeled in the
limit of negligible accretion and it is the most direct
manifestation of gravitohydromagnetics (GHM) \citet{pun01}. A GHM
dynamo arises when the magnetic field impedes the inflow of gas in
the ergosphere, i.e., vertical flux in an equatorial accretion flow.
The strong gravitational force will impart stress to the magnetic
field in an effort to move the plasma through the obstructing flux.
In particular, the metric induced frame dragging force will twist up
the field azimuthally. These stresses are coupled into the accretion
vortex around a black hole by large scale magnetic flux, and
propagate outward as a relativistic Poynting jet. The more obstinate
the obstruction, the more powerful the jet. There are two defining
characteristics that distinguish the GHM dynamo from a
Blandford-Znajek (B-Z) process, \citet{blz77}, on field lines that
thread the ergopshere:
\begin{enumerate}
\item The B-Z process is electrodynamic so there is no source within
the ergosphere, it appears as if the energy flux is emerging from
the horizon. In the GHM mechanism, the source of Poynting flux is in
the ergospheric equatorial accretion flow.
\item In a B-Z process in a magnetosphere shaped by the accretion vortex,
the field line angular velocity is, $\Omega_{F}\approx \Omega_{H}/2$
(where $\Omega_{H}$ is the angular velocity of the horizon) near the
pole and decreases with latitude to $\approx \Omega_{H}/5$ near the
equatorial plane of the inner ergosphere \citep{phi83}. In GHM,
since the magnetic flux is anchored by the inertia of the accretion
flow in the inner ergosphere, frame dragging enforces $d \phi/dt
\approx \Omega_{H}$. One therefore has the condition,
$\Omega_{F}\approx \Omega_{H}$.

\end{enumerate}
In order to understand the physical origin of the Poynting flux,
these two issues are studied below.

\begin{figure}
\epsscale{0.95} \plottwo{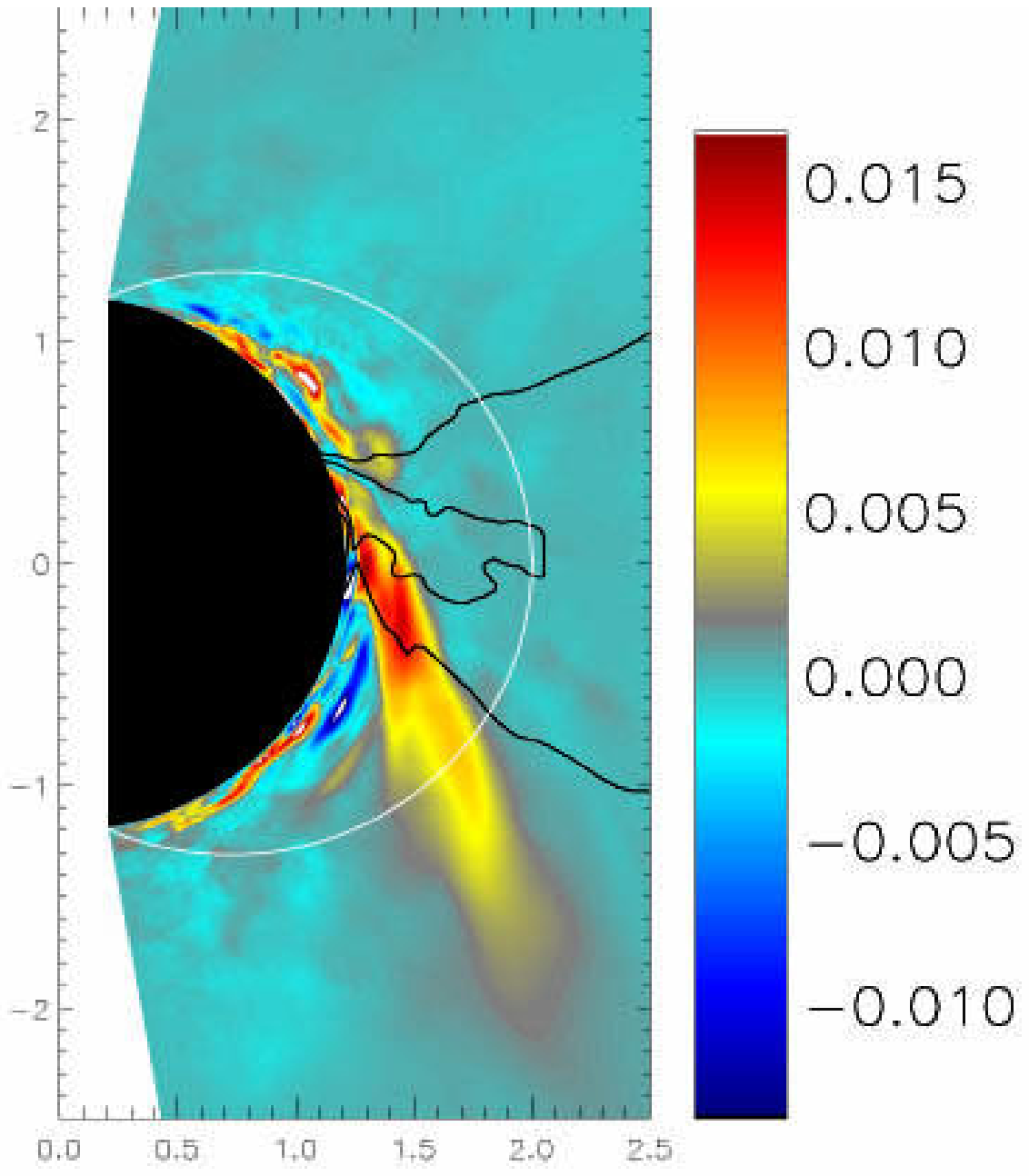}{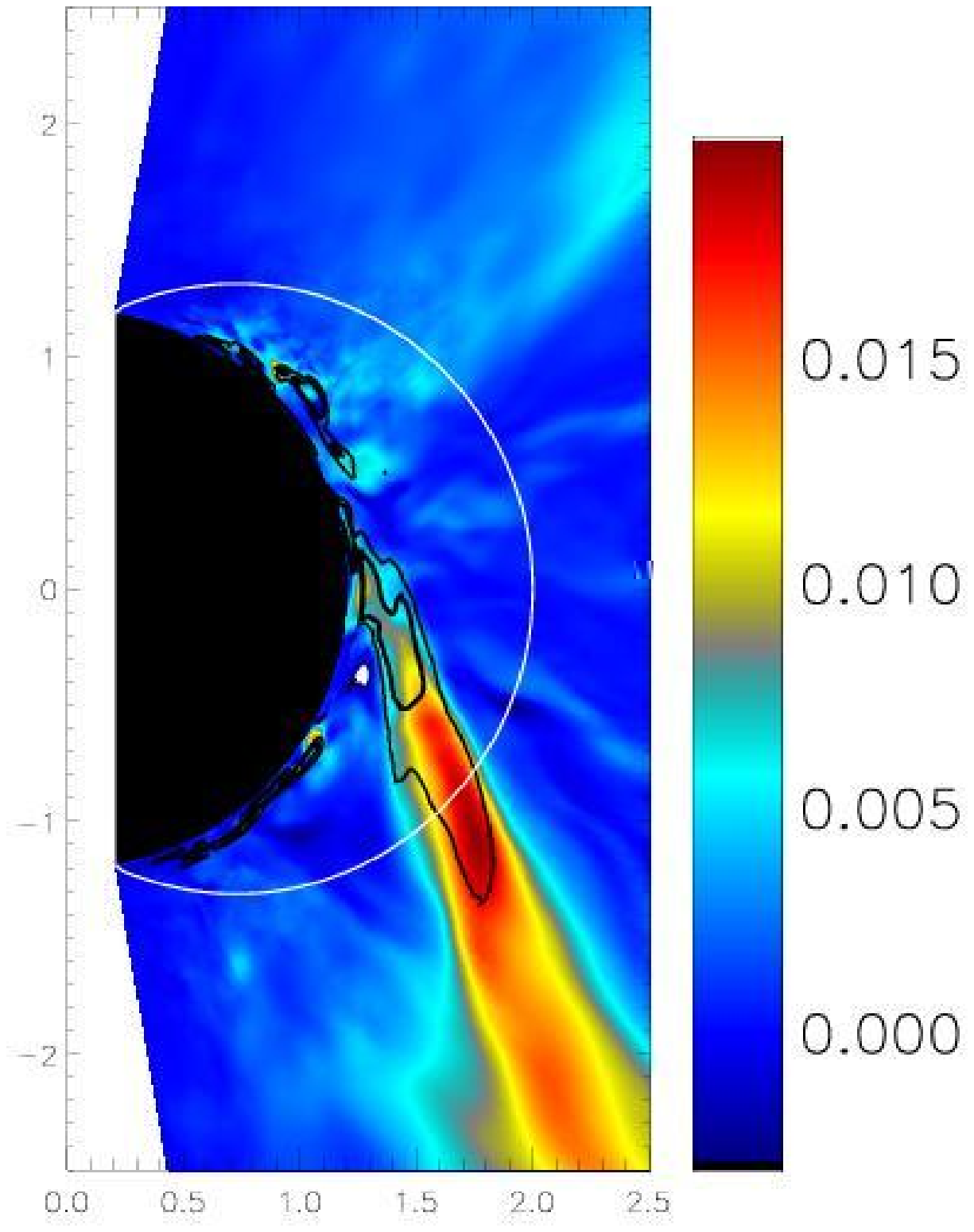}\\
\epsscale{0.95} \plottwo{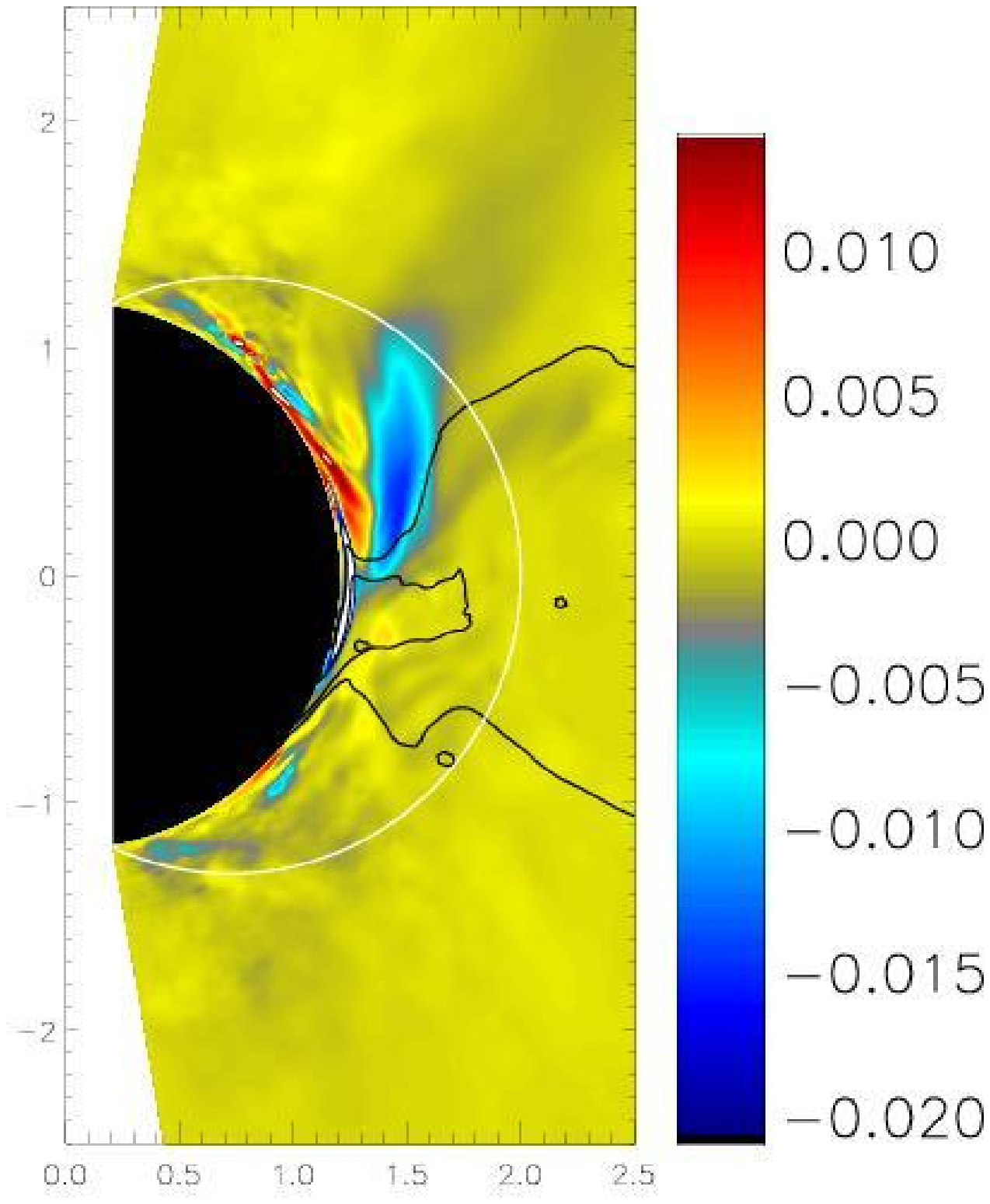}{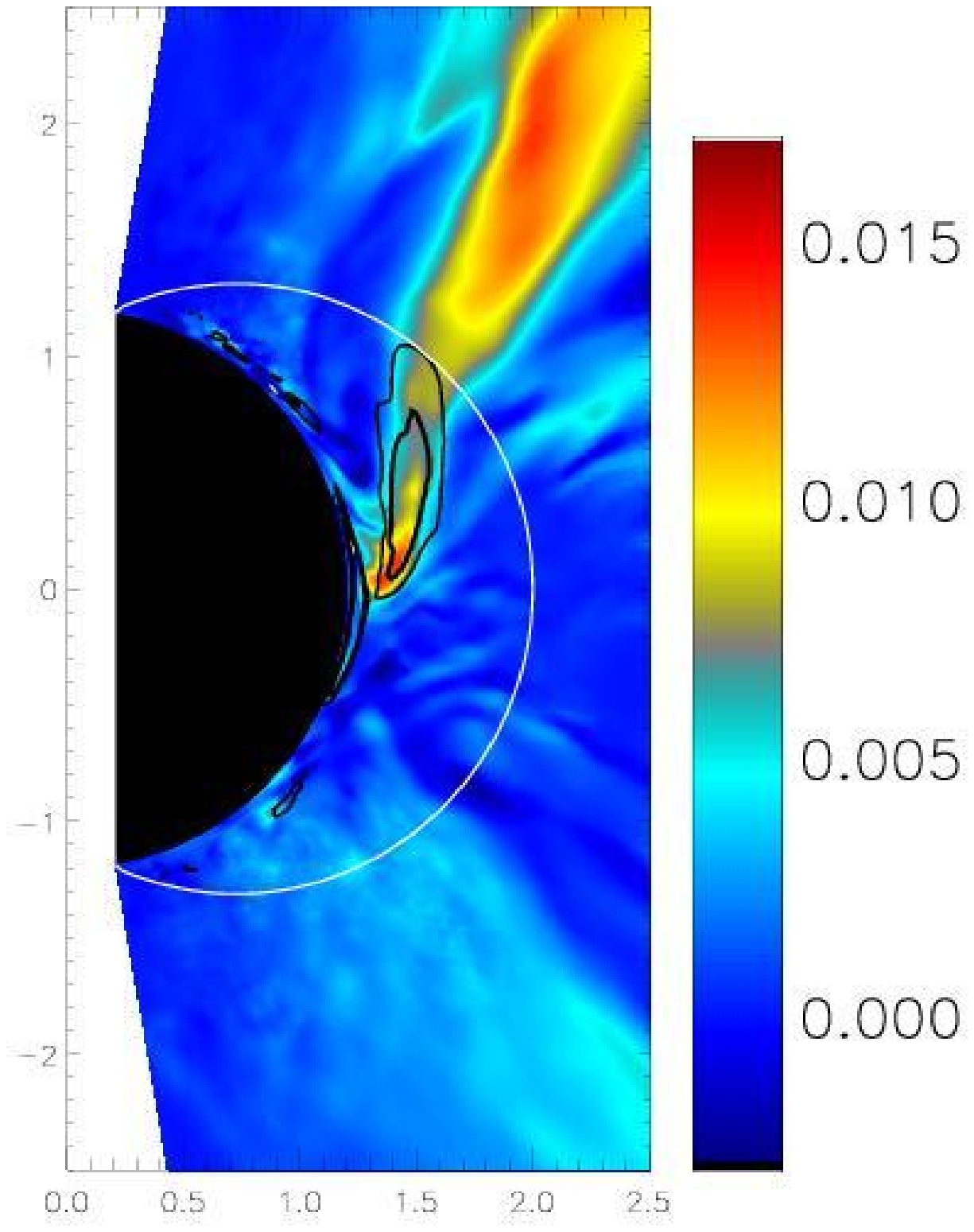}\\
\epsscale{0.95}\plottwo{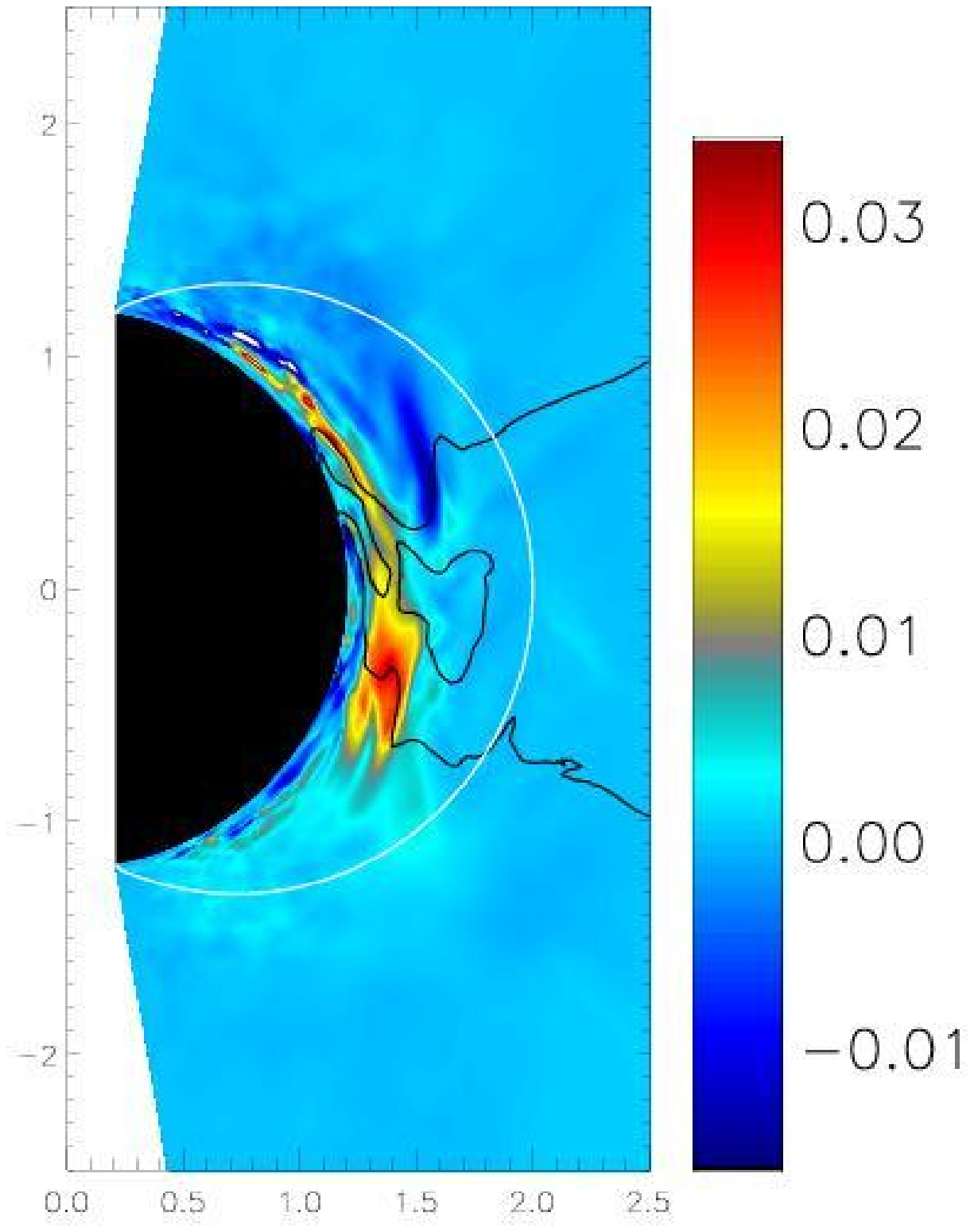}{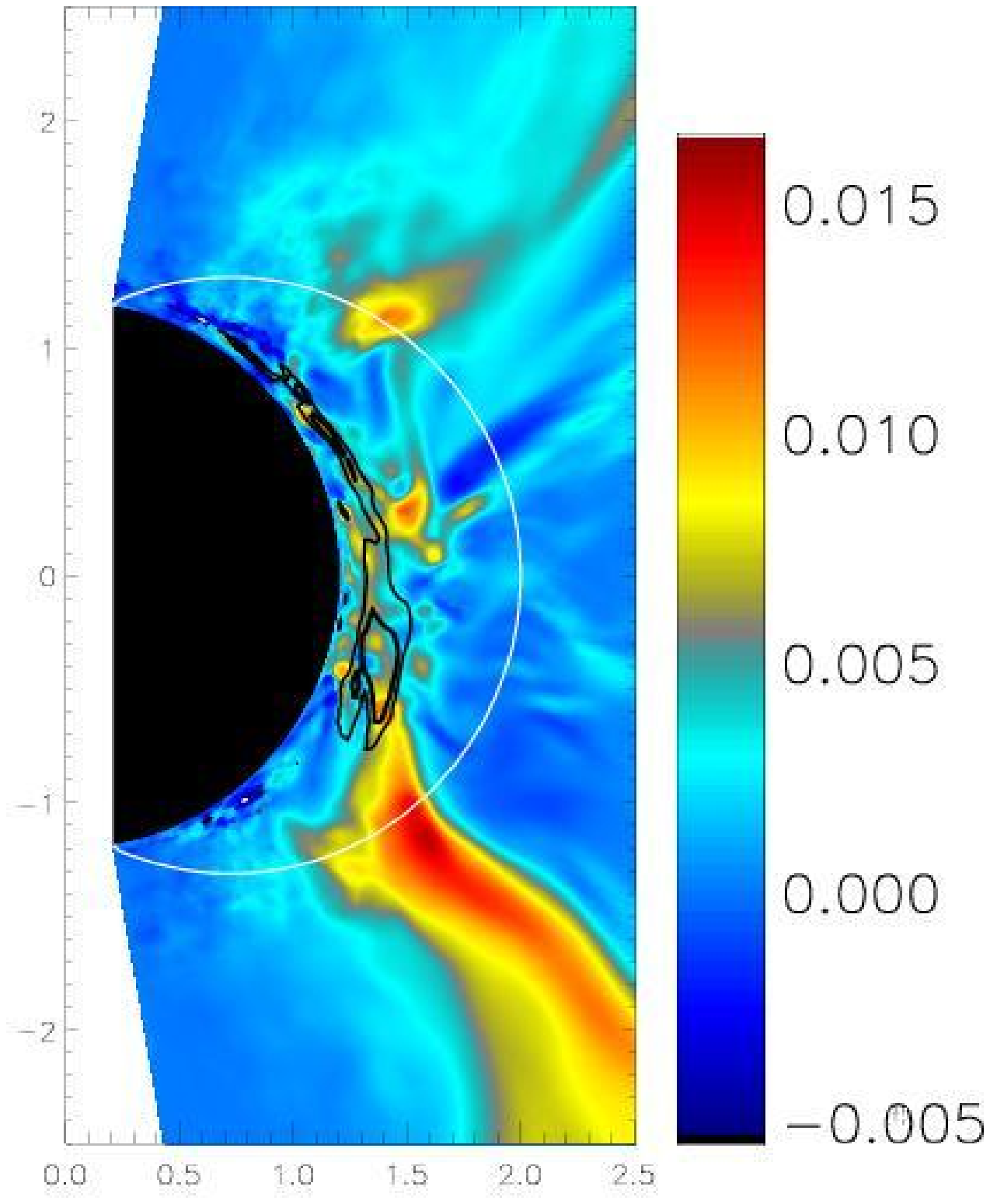}
\caption{The source of Poynting flux. The
left hand column is $S^{\theta}$ and the right hand column is
$S^{r}$ in KDJ, both averaged over azimuth, at (from top to bottom)
t= 9840 M, t = 9920 M and t= 10000 M. The relative units (based on
code variables) are in a color bar to right of each plot for
comparison of magnitudes between the six plots. The contours on the
$S^{\theta}$ plots are of the density, scaled from the peak value
within the frame at relative levels 0.5 and 0.1. The contours on the
$S^{r}$ plots are of $S^{\theta}$ scaled from the peak within the
frame at relative levels 0.67 and 0.33. The inside of the inner
calculational boundary (r=1.203 M) is black. The calculational
boundary near the poles is at $8.1^{\circ}$ and $171.9^{\circ}$.
Notice that any contribution from an electrodynamic effect
associated with the horizon appears minimal. The white contour is
the stationary limit surface. There is no data clipping, so plot
values that exceed the limits of the color bar appear white.}
\end{figure}

\begin{figure}
\epsscale{0.95} \plottwo{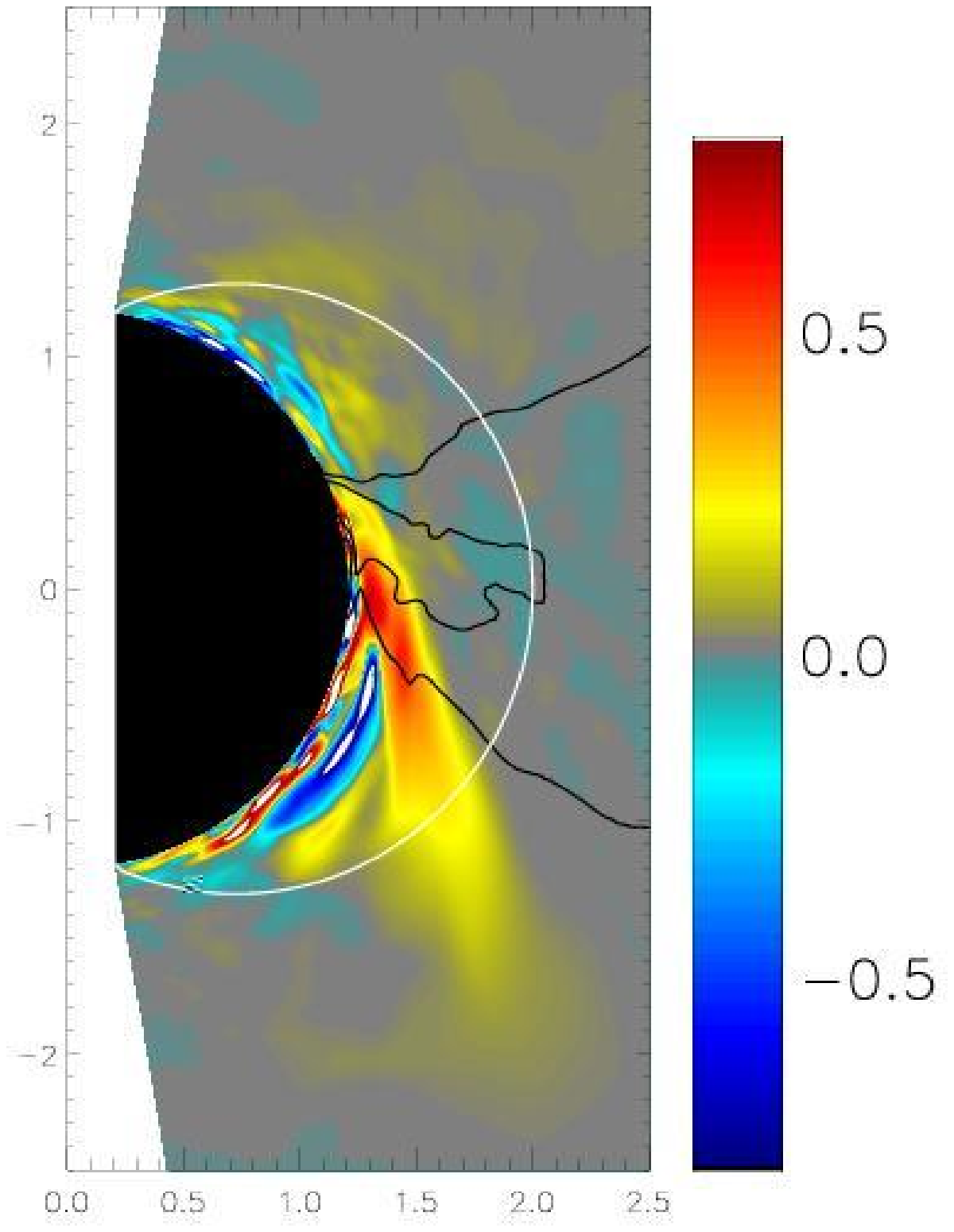}{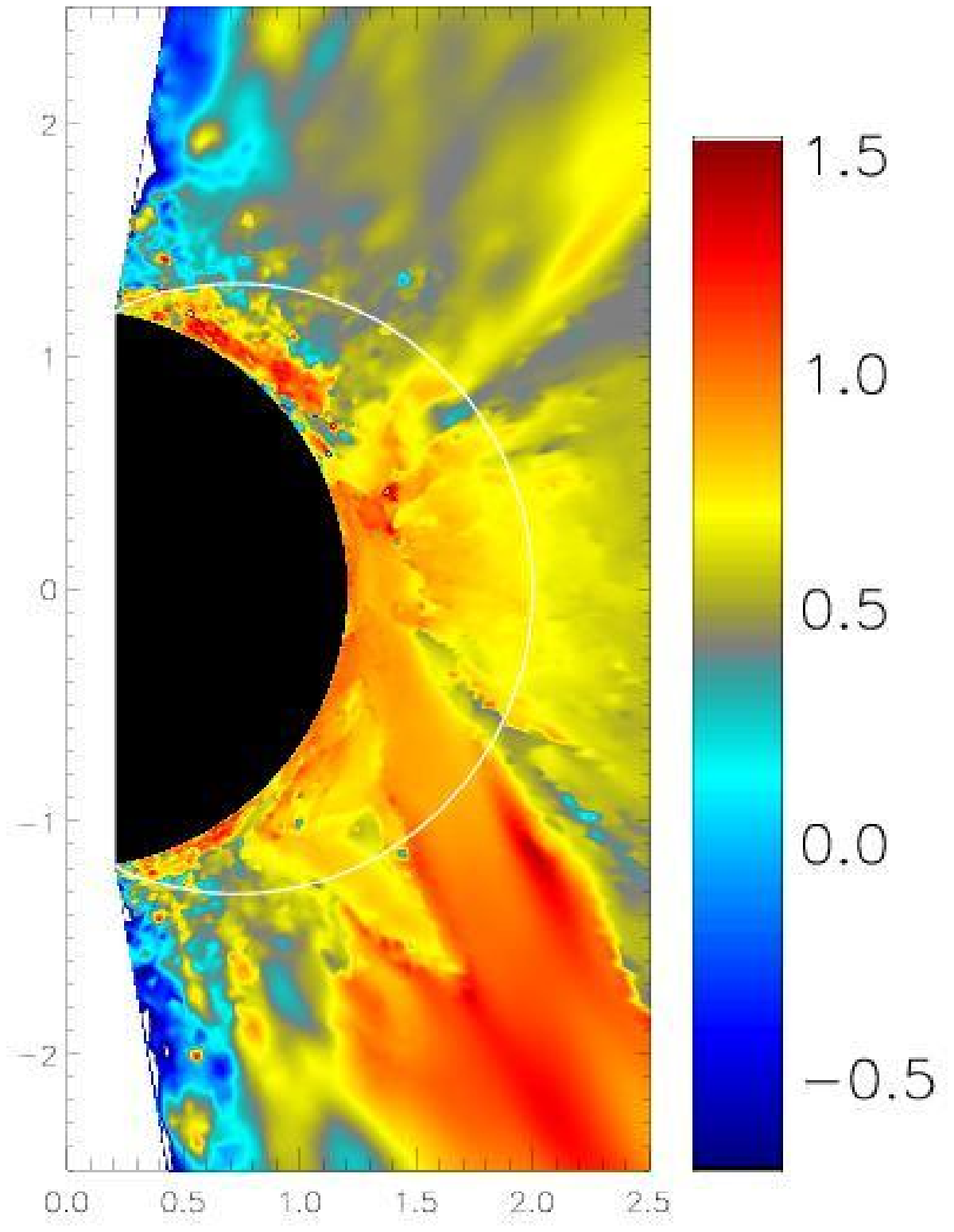}\\
\epsscale{0.95} \plottwo{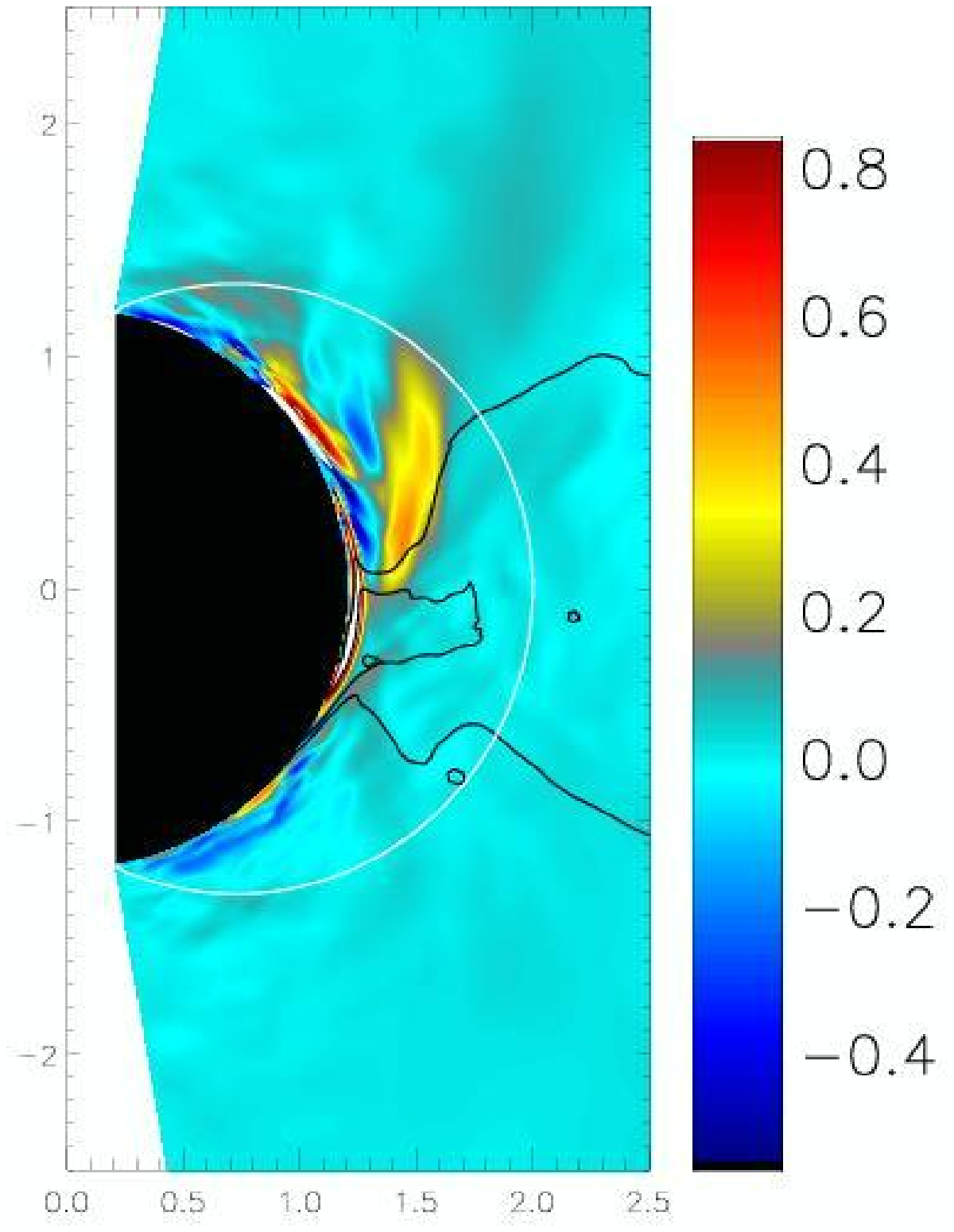}{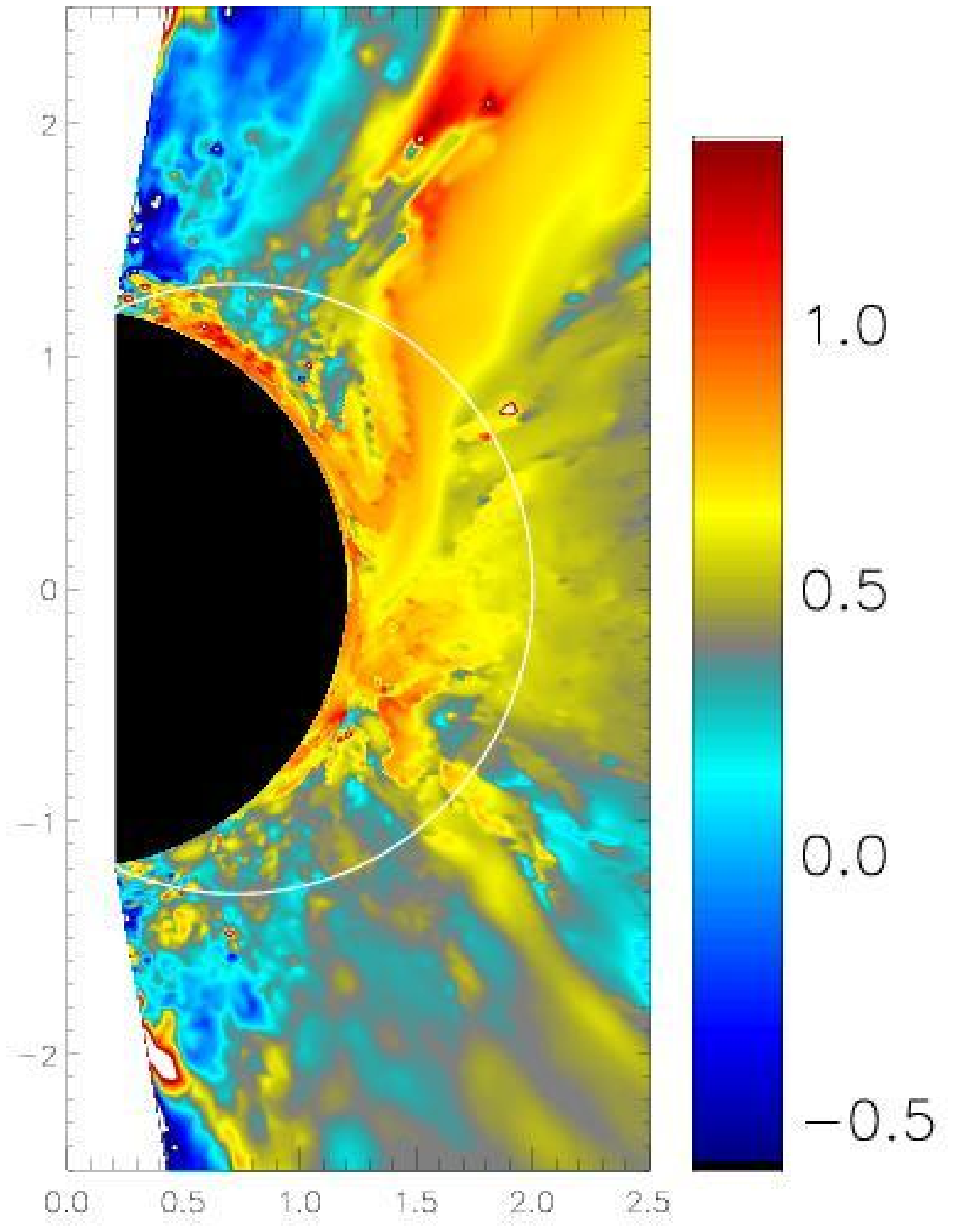}\\
\epsscale{0.95} \plottwo{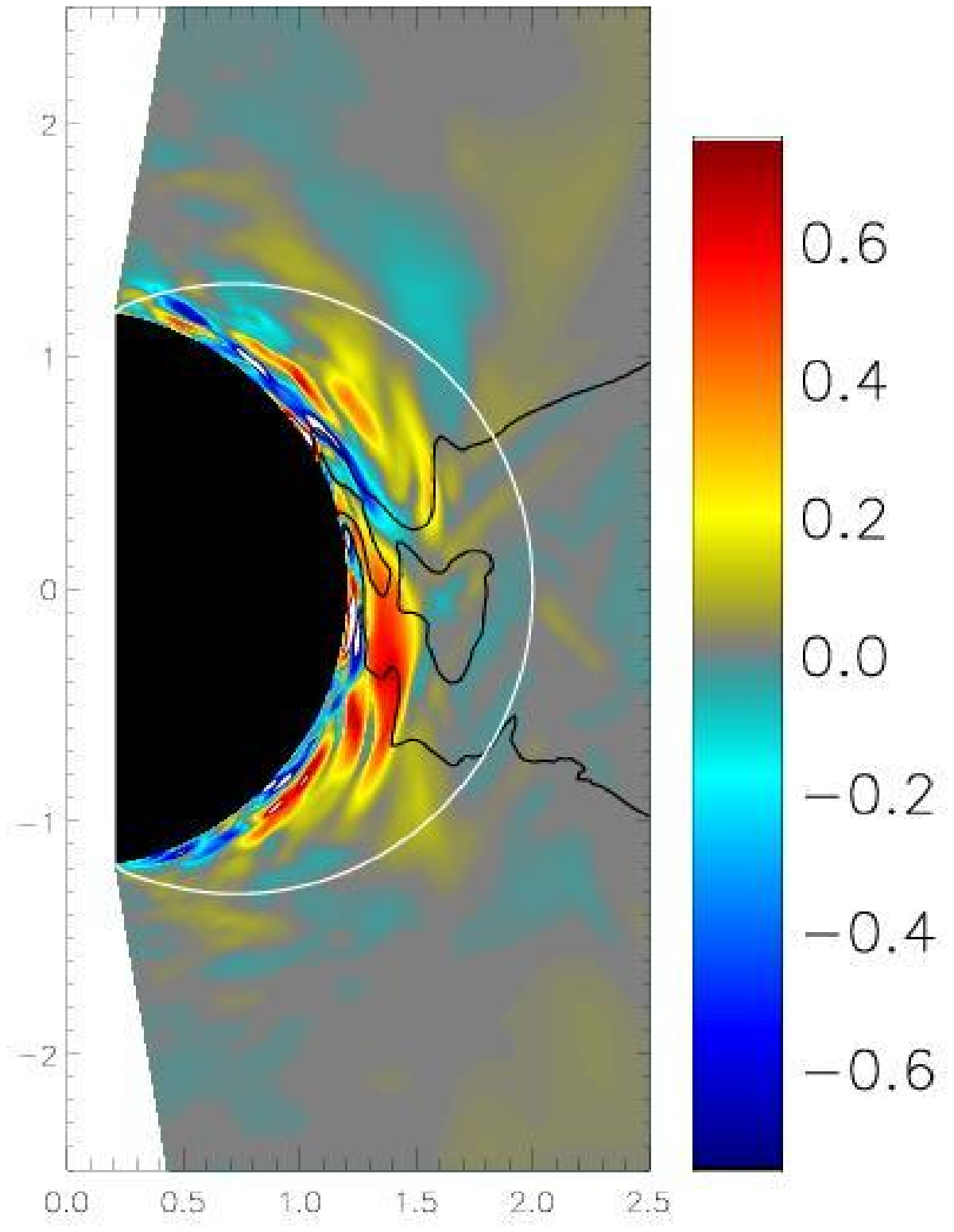}{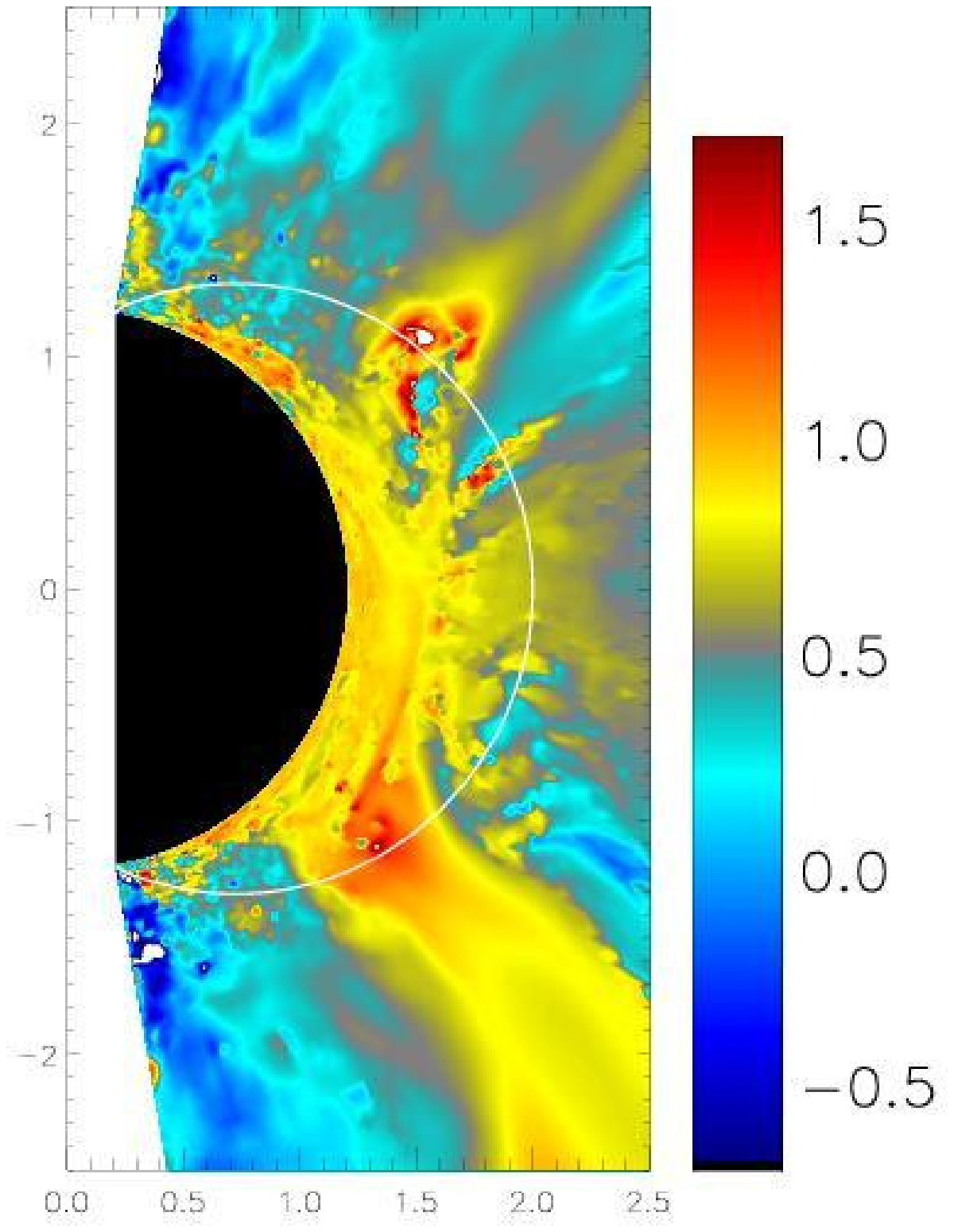}
\caption{The central engine. The left hand column is $B^{\theta}$ and the right hand column is
$\Omega_{F}$ in KDJ, both averaged over azimuth, at (from top to
bottom) t= 9840 M, t = 9920 M and t= 10000 M. The relative units
(based on code variables) are in a color bar to right of each plot
for comparison of magnitudes between the plots. The calculational
boundaries are the same as figure 1. The contours on the
$B^{\theta}$ plots are of the density, scaled from the peak value
within the frame at relative levels 0.5 and 0.1. There is no data
clipping, so plot values that exceed the limits of the color bar
appear white.}
\end{figure}

\section{The KDJ Simulation}
The simulation is performed in the Kerr metric (that of a rotating,
uncharged black hole), $g_{\mu\nu}$. Calculations are carried out in
Boyer-Lindquist (B-L) coordinates $(r,\theta,\phi,t)$. The reader
should refer to \citet{haw06} for details of the simulation. We only
give a brief overview. The initial state is a torus of gas in
equilibrium that is threaded by concentric loops of weak magnetic
flux that foliate the surfaces of constant pressure. The magnetic
loops are twisted azimuthally by the differentially rotating gas.
This creates significant magnetic stress that removes angular
momentum from the gas, initiating a strong inflow that is permeated
by magneto-rotational instabilities (MRI). The end result is that
after t = a few hundred M, accreted poloidal magnetic flux gets
trapped in the accretion vortex or funnel (with an opening angle of
$\sim 60^{\circ}$ at the horizon tapering to $ \sim 35^{\circ}$ at
$r > 20 M $). This region is the black hole magnetosphere and it
supports a Poynting jet. The surrounding accretion flow is very
turbulent.
\par In order to understand the source of the strong flares of radial
Poynting flux, one needs to merely consider the conservation of
global, redshifted, or equivalently the B-L coordinate evaluated
energy flux \citep{thp86}. In general, the divergence of the time
component of the stress-energy tensor in a coordinate system can be
expanded as, $T_{t\; ;\nu}^{\; \nu}=
(1/\sqrt{-g})[\partial(\sqrt{-g}\, T_{t}^{\,
\nu})/\partial(x^{\nu})] + \Gamma^{\mu}_{\; t \; \beta} T_{\mu}^{\,
\beta}$, where $\Gamma^{\mu}_{\; t \; \beta}$ is the connection
coefficient and $g = -(r^{2} + a^{2}
\cos^{2}{\theta})^{2}\sin^{2}{\theta}$ is the determinant of the
metric. However, the Kerr metric has a Killing vector (the metric is
time stationary) dual to the B-L time coordinate. Thus, there is a
conservation law associated with the time component of the
divergence of the stress-energy tensor. Consequently, if one expands
out the inhomogeneous connection coefficient term in the expression
above, it will equate to zero. The conservation of energy evaluated
in B-L coordinates reduces to, $\partial(\sqrt{-g}\, T_{t}^{\,
\nu})/\partial(x^{\nu})=0 $, where the four-momentum $-T_{t}^{\,
\nu}$ has two components: one from the fluid, $-(T_{t}^{\,
\nu})_{\mathrm{fluid}}$, and one from the electromagnetic field,
$-(T_{t}^{\, \nu})_{\mathrm{EM}}$. The reduction to a homogeneous
equation with only partial derivatives is the reason why the global
conservation of energy can be expressed in integral form in (3.70)
of \citet{thp86}. It follows that the poloidal components of the
redshifted Poynting flux are $S^{\theta}= -\sqrt{-g}\, (T_{t}^{\,
\theta})_{\mathrm{EM}}$ and $S^{r}= -\sqrt{-g}\, (T_{t}^{\,
r})_{\mathrm{EM}}$. We can use these simple expressions to
understand the primary source of the Poynting jet in KDJ. Figure 1
is a plot of $S^{\theta}$ (on the left) and $S^{r}$ (on the right)
in KDJ at the last three time steps of data collection. Each frame
is the average over azimuth of each time step. This greatly reduces
the fluctuations as the accretion vortex is a cauldron of strong MHD
waves. The individual $\phi=\mathrm{constant}$ slices show the same
dominant behavior, however it is embedded in large MHD fluctuations.
On the left hand column of figure 1, density contours have been
superimposed on the images to indicate the location of the
equatorial accretion flow. The density is evaluated in B-L
coordinates with contours at 0.5 and 0.1 of the peak value within $r
< 2.5M$. Notice that in all three left hand frames, $S^{\theta}$ is
created primarily in regions of very high accretion flow density. In
all three of the right hand frames of figure 1, there is an enhanced
$S^{r}$ that emanates from the ergosphere (defined by the interior
of the stationary limit, $r_{s} = M + \sqrt{M^{2}
-a^{2}\cos^{2}{\theta}}$, note that there are 40 grid points between
$r=1.203 M$ and $r_{s}$ at $\theta =\pi/2$). This radial energy beam
diminishes precipitously just outside the horizon, near the
equatorial plane in all three time steps. The region in which
$S^{r}$ diminishes is adjacent to a region of strong $S^{\theta}$
that originates in the inertially dominated accretion flow in the
inner ergosphere, $1.2 M < r < 1.6 M $ (this region is resolved by
28 radial grid zones). In fact, if one looks at the conservation of
energy equation, the term $\partial (S^{\theta})/\partial {\theta}$
is sufficiently large to be the source of $\partial (S^{r})/\partial
{r}$ at the base of the radial beam in all three frames. This does
not preclude the transfer of energy to and from the plasma. It
merely states that the magnitude is sufficient to source $S^{r}$. In
general, the hydrodynamic energy flux is negligible in the funnel.
In order to illustrate this, contours of $S^{\theta}$ are
superimposed on the color plots of $S^{r}$. The contour levels are
chosen to be 2/3 and 1/3 of the maximum value of $S^{\theta}$
emerging from the dense equatorial accretion flow. One clearly sees
$S^{\theta}$ switching off where $S^{r}$ switches on. We conclude
that a vertical Poynting flux created in the equatorial accretion
flow is the source of the strong beams of $S^{r}$. This establishes
condition 1 of the Introduction.
\par The left column of
figure 2 contains plots of the magnetic field component,
$B^{\theta}\equiv F_{r \phi}$, at the three time steps. At every
location in which $S^{\theta}$ is strong in figure 1, there is a
pronounced enhancement in $B^{\theta}$ in figure 2. Recall that the
sign of $S^{\theta}$ is not determined by the sign of $B^{\theta}$.
These intense flux patches penetrate the inertially dominated
equatorial accretion flow in all three frames. The density contours
indicate that the regions of enhanced vertical field greatly disrupt
the equatorial inflow. As noted in the introduction, a GHM
interaction is likely to occur when the magnetic field impedes the
inflow in the ergosphere. The regions of large $B^{\theta}$ are
compact compared to the global field configuration of the jet, only
$\sim 1.0 M - 2.0 M$ long. Considering the turbulent, differentially
rotating plasma in which they are embedded, these are most likely
highly enhanced regions of twisted magnetic loops created by the
MRI. The strength of $B^{\theta}$ at the base of the flares is
comparable to, or exceeds the radial magnetic field strength. The
situation is clearly very unsteady and vertical flux is constantly
shifting from hemisphere to hemisphere. The time slice t = 10000 M,
although primarily a southern hemisphere event, also has a
significant contribution in the northern hemisphere (see the blue
fan-like plume of vertical Poynting flux in figure 1). The GHM
interaction is provided by the vertical flux that links the
equatorial plasma to the relatively slowly rotating plasma of the
magnetosphere within the accretion vortex. The vertical flux
transmits huge torsional stresses from the accretion flow to the
magnetosphere.
\par Further corroboration of this interpretation
can be found by looking at the values of $\Omega_{F}$ in the
vicinity of the $S^{r}$ flares. In a non-axisymmetric, non-time
stationary flow, there is still a well defined notion of
$\Omega_{F}$: the rate at which a frame of reference at fixed r and
$\theta$ would have to rotate so that the poloidal component of the
electric field, $E^{\perp}$, that is orthogonal to the poloidal
magnetic field, $B^{P}$, vanishes. This was first derived in
\citet{pun91} (see the extended discussion in \citet{pun01} for the
various physical interpretations), and has recently been written out
in B-L coordinates in \citet{haw06} in terms of the plasma
three-velocity, $v^{i}$ and the Faraday tensor as
\begin{eqnarray}
&& \Omega_{F} = v^{\phi} - F_{\theta r}\frac{g_{rr}v^{r}F_{\phi
\theta} + g_{\theta \theta} v^{\theta} F_{r \phi}}{(F_{\phi
\theta})^{2} g_{rr} + (F_{r \phi})^{2} g_{\theta \theta}}\;.
\end{eqnarray}
This expression was studied in the context of the simulation KDH,
$a/M=0.95$, in \citet{haw06}. They found that a long term time and
azimuth average yielded $\Omega_{F}\approx 1/3\Omega_{H} $ and there
was no enhancement at high latitudes as was anticipated by
\citet{phi83}. The t = 10000 M time slice of KDH was generously
provided to this author. At t = 10000 M, there are no strong flares
emerging from the equatorial accretion flow. Inside the funnel at $r
< 10M$, at t=10000 M, $0<\Omega_{F} < 0.5\Omega_{H}$.
\par The right hand column of figure 2 is
$\Omega_{F}$ plotted at three different time steps for KDJ. By
comparison to figure 1, notice that each flare in $S^{r}$ is
enveloped by a region of enhanced $\Omega_{F}$, typically $0.7
\Omega_{H}< \Omega_{F} < 1.2 \Omega_{H}$. The regions of the funnel
outside the ergosphere are devoid of large flares in $S^{r}$ and
typically have $0<\Omega_{F} < 0.5\Omega_{H}$, similar to what is
seen in KDH.. Unlike KDH, there are huge enhancements in
$\Omega_{F}$ at lower latitudes in the funnel. It seems reasonable
to associate this large difference in the peak values of
$\Omega_{F}$ in KDJ and KDH (at t= 10000 M) with the spatially and
temporally coincident flares in $S^{r}$ that occur in KDJ.
Furthermore, this greatly enhanced value of $\Omega_{F}$ indicates a
different physical origin for $\Omega_{F}$ in the flares than for
the remainder of the funnel or in KDH at t = 10000 M. The most
straightforward interpretation is that it is a direct consequence of
the fact that the flares originate on magnetic flux that is locked
into approximate corotation with the dense accreting equatorial
plasma (i.e., the inertially dominated equatorial plasma anchors the
magnetic flux). In the inner ergosphere, frame dragging enforces $
0.7\Omega_{H} < d\phi/ dt < 1.0 \Omega_{H}$ on the accretion flow.
This establishes condition 2 of the Introduction.
\section{Discussion}In this Letter we showed that in the last three data dumps of
the 3-D MHD numerical simulation, KDJ, the dominant source of
Poynting flux originated near the equatorial plane deep in the
ergopshere. The phenomenon is unsteady and is triggered by large
scale vertical flux that is anchored in the inertially dominated
equatorial accretion flow. The situation typifies the ergospheric
disk in virtually every aspect, even though there is an intense
accretion flow. There is one exception, unlike the ergospheric disk,
the anchoring plasma rarely achieves the global negative energy
condition that is defined by the four-velocity, $-U_{t}<0$, because
of the flood of incoming positive energy plasma from the accretion
flow. The plasma attains $-U_{t}<0$ only near the base of the
strongest flares seen in the $\phi=\mathrm{constant}$ slices.
\par The switch-on of a powerful
beam of $S^{r}$ outside the horizon at $r\approx 1.3M $ in the
$a/M=0.995$ simulation, KDE, of \citet{kro05} was demonstrated in
\citet{pun05}. It seems likely the the source of $S^{r}$ in KDE is
$S^{\theta}$ from an ergopsheric disk. The ergospheric disk appears
to switch on at $a/M>0.95$ as evidenced by the factor of 3 weaker
Poynting flux in KDH. Furthermore, if the funnel opening angle at
the horizon in KDH at t= 10000 M is typical within $\pm 5^{\circ}$
then figure 5 and table 4 of \citet{haw06} indicate that only 35\%
to 40\% of the funnel Poynting flux at large distances is created
outside the horizon during the course of the simulation. A plausible
reason is given by the plots of $B^{\theta}$ in figure 2. The
vertical magnetic flux at the equatorial plane is located at $r <
1.55 M$. The power in the ergospheric disk jet $\sim
[B^{\theta}(SA)(\Omega_{H})]^{2}$, where SA is the proper surface
area of the equatorial plane threaded by vertical magnetic flux
\citep{sem04,pun01}. The proper surface area in the ergospheric
equatorial plane increases dramatically at high spin, diverging at
$a=M$. For example, between the inner calculational boundary and
1.55 M the surface area is only significant for $a/M>0.95$ and grows
quickly with $a/M$, exceeding twice the surface area of the horizon
for $a/M=0.99$. Thus, if $B^{\theta}$ in the inner ergosphere were
independent of spin to first order, then a strong ergospheric disk
jet would switch-on in the 3-D simulations at $a/M>0.95$. Note that
if the inner boundary were truly the event horizon instead of the
inner calculational boundary then this argument would indicate that
the ergospheric disk would likely be very powerful even at
$a/M=0.95$ and the switch-on would occur at $a/M \approx 0.9$. The
implication is that a significant amount of large scale magnetic
flux threading the equatorial plane of the ergopshere (which implies
a large black hole spin based on geometrical considerations)
catalyzes the formation of the most powerful Poynting jets around
black holes. Thus, we are now considering initial conditions in
simulations that are conducive to producing significant vertical
flux in the equatorial plane of the ergosphere.
\par It should be noted that 2-D simulations from a similar initial
state of torii threaded by magnetic loops have been studied in
\citet{gam04}. However, the magnetic flux evolution can be much
different in this setting as discussed in \citet{pun05} and poloidal
flux configurations conducive to GHM could be highly suppressed. In
summary, there are no interchange instabilities, so flux tubes
cannot pass by each other or move around each other in the extra
degree of freedom provided by the azimuth. Thus, there is a tendency
for flux tubes to get pushed into the hole by the accretion flow.
This is in contrast to the formation of the ergospheric disk in
\citet{pun90} in which buoyant flux tubes are created by
reconnection at the inner edge of the ergospheric disk and recycle
back out into the outer ergosphere by interchange instabilities.
Ideally, a full 3-D simulation with a detailed treatment of
resistive MHD reconnection is preferred for studying the relevant
GHM physics.
\begin{acknowledgements}
I would like to thank Jean-Pierre DeVilliers for sharing his deep
understanding of the numerical code and these simulations. I was
also very fortunate that Julian Krolik and John Hawley were willing
to share their data in the best spirit of science.
\end{acknowledgements}

\end{document}